\documentclass[a4paper,fleqn,usenatbib]{mnras}

\usepackage[T1]{fontenc}
\usepackage{ae,aecompl}

\usepackage{subcaption}
\usepackage{graphicx}	
\usepackage{amsmath}	
\usepackage{amssymb}	

\title[Investigating spectral features of SNRs with SRT]
{Investigating the high-frequency spectral features of SNRs Tycho, W44 and IC443 with the Sardinia Radio Telescope}

\author[S. Loru et al.]{
S. Loru,$^{1,2}$\thanks{E-mail: saraloru@oa-cagliari.inaf.it}
A. Pellizzoni$^{1}$,
E. Egron$^{1}$,
S. Righini$^{3}$,
M. N. Iacolina$^{4}$,
S. Mulas$^{2}$,
\newauthor
M. Cardillo$^{5}$,
M. Marongiu$^{6,7}$,
R. Ricci$^{3}$,
M. Bachetti$^{1}$,
M. Pilia$^{1}$,
A. Trois$^{1}$,
\newauthor
A. Ingallinera$^{8}$,
O. Petruk$^{9}$,
G. Murtas$^{2}$,
G. Serra$^{1}$,
F. Buffa$^{1}$,
R. Concu$^{1}$,
\newauthor
F. Gaudiomonte$^{1}$,
A. Melis$^{1}$,
A. Navarrini$^{1}$,
D. Perrodin$^{1}$,
G. Valente$^{4}$
\\
$^{1}$INAF, Osservatorio Astronomico di Cagliari, Via della Scienza 5, 09047 Selargius, Italy\\
$^{2}$Dipartimento di Fisica, Universit{\'a} degli Studi di Cagliari, SP Monserrato-Sestu, KM 0.7, 09042 Monserrato, Italy\\
$^{3}$INAF, Istituto di Radio Astronomia di Bologna, Via P. Gobetti 101, 40129 Bologna, Italy\\
$^{4}$ASI,Osservatorio Astronomico di Cagliari, Via della Scienza 5, 09047 Selargius, Italy\\
$^{5}$INAF, Istituto di Astrofisica e Planetologia Spaziali,Via Fosso del Cavaliere 100, 00133 Roma, Italy\\
$^{6}$ Department of Physics and Earth Sciences, University of Ferrara, via Saragat 1, 44122 Ferrara, Italy\\
$^{7}$ ICRANet, Piazzale della Repubblica 10, 65122, Pescara, Italy\\
$^{8}$INAF, Osservatorio Astrofisico di Catania, Via Santa Sofia 78, I-95123 Catania, Italy\\
$^{9}$Institute for Applied Problems in Mechanics and Mathematics, Naukova Str. 3-b, UA-79060 Lviv, Ukraine\\
}

\date{Accepted XXX. Received YYY; in original form ZZZ}

\pubyear{}

\begin{document}
\label{firstpage}
\pagerange{\pageref{firstpage}--\pageref{lastpage}}
\maketitle

\begin{abstract}
The main characteristics in the radio continuum spectra of Supernova Remnants (SNRs) result from simple synchrotron emission. 
In addition, electron acceleration mechanisms can shape the spectra in specific ways, especially at high radio frequencies. These features are connected to the age and the peculiar conditions of the local interstellar medium interacting with the SNR. 
Whereas the bulk radio emission is expected at up to $20-50$ GHz, sensitive high-resolution images of SNRs above 10 GHz are lacking and are not easily achievable, especially in the confused regions of the Galactic Plane.
In the framework of the early science observations with the Sardinia Radio Telescope in February-March 2016, we obtained high-resolution images of SNRs Tycho, W44 and IC443
that provided accurate integrated flux density measurements at 21.4 GHz: 8.8 $\pm$ 0.9 Jy for Tycho, 25 $\pm$ 3 Jy for W44 and 66 $\pm$ 7 Jy for IC443.
We coupled the SRT measurements with radio data available in the literature in order to characterise the integrated and spatially-resolved spectra of these SNRs, and to find significant frequency-  and region-dependent spectral slope variations.
For the first time, we provide direct evidence of a spectral break in the radio spectral energy distribution of W44 at an exponential cutoff frequency of 15 $\pm$ 2 GHz. This result constrains the maximum energy of the accelerated electrons in the range $6-13$ GeV, in agreement with predictions indirectly derived from  AGILE and \textit{Fermi}-LAT gamma-ray observations. 
With regard to IC443, our results confirm the noticeable presence of a bump in the integrated spectrum around $20-70$ GHz that could result from a spinning dust emission mechanism.

\end{abstract}



\newpage
\section{Introduction}

Supernova remnants (SNRs) result from the stellar mass ejected during a supernova explosion, and their spatial and temporal evolution is strongly related to the interaction between the SNR shocks and the surrounding interstellar medium (ISM).
The expanding ejecta carry a huge amount of energy ($\sim$10$^{50}$ erg) that is transferred to the magnetic field and kinetic/thermal energy of the shocked interstellar gas and relativistic particles.\\
Electrons and heavier particles are responsible for the non-thermal emissions of SNRs, both in the radio and gamma-ray bands. 
In the radio band, the spectral energy distribution (SED) of SNRs is characterised by synchrotron emission ($S_{\nu}\propto \nu^{-\alpha}$ with $\alpha \sim 0.5$) from relativistic electrons accelerated into the magnetic field by a diffusive shock mechanism.
Assuming a magnetic field of the order of 10 $\mu$G, which typically arises by  compression  of the ISM in SNR shocks \citep{Wielebinski_2005}, it is expected that 0.5$-$50 GeV electrons emit the observed synchrotron radiation in the  0.01$-$100 GHz frequency range. 
A steepening in the radio spectrum is associated with a decrease in the energy distribution of the emitting electrons.\\
The gamma-ray emission results from both hadronic ($\pi^0$ mesons decay) and leptonic (Bremsstrahlung and Inverse Compton, IC) processes. The latter show bumps that are bound to the synchrotron spectral slope and to the cutoff in the radio domain. 
From the gamma-ray observations of middle-aged (with ages greater than a few thousand years) SNRs like W44 and IC443, which showed a steepening of the primary particle spectrum at energies of 10$-$100 GeV, a synchrotron cutoff is expected above 10 GHz \citep{Ackermann_2013}. \\
It is therefore of great interest to directly measure the high-frequency radio spectral breaks in order to model the multi-wavelength SED of SNRs.
This could provide crucial information for disentangling the two basic particle acceleration processes represented at high energies by leptonic and hadronic models, and for better constraining SNRs as cosmic ray (CR) emitters (\citealt{Ackermann_2013}, \citealt{Cardillo_2014},  \citealt{Cardillo_2016}). \\
Furthermore, the co-spatial study of the radio and gamma-ray emissions coupled with a spatially-resolved analysis are required in order to discern possible radio emission mechanisms related to electron populations at different evolutionary stages, or located in SNR regions with different ambient conditions (inhomogeneous medium, molecular clouds interactions or different shock conditions).\\  
Despite this promising perspective, the shape of the radio continuum spectra of Galactic SNRs at high frequencies is far from being fully understood due to a lack of accurate integrated flux
densities. Indeed, sensitive radio flux density measurements are typically available in the literature only at up to 5$-$10 GHz for large SNRs with a complex morphology.
Interferometric observations provide significant amount of information about the morphology of SNRs, but accurate flux estimation and imaging of large structures of $\sim$ 10$-$30 arcmin is unfeasible above a few GHz.
Observations of Galactic SNRs at up to microwave frequencies have been carried out by the $Planck$ and QUIJOTE instruments (\citealt{Planck_2016}, \citealt{Genova-Santos_2017}) with a relatively low resolution that makes it difficult to provide precise flux densities of sources located in crowded regions of the Galactic Plane. 
On the other hand, sensitive radio continuum observations can be performed with single-dish telescopes, offering in this way a good compromise between sensitivity and resolution in the frequency range 5$-$50 GHz.\\
Among nearby and large sources, SNRs W44 and IC443 are the best suited for multi-wavelength spectral studies ranging from radio to gamma-rays \citep{Ackermann_2013}. That is because these ``mixed-morphology'' sources offer a rich region-dependent phenomenology including atomic/molecular clouds interactions, Pulsar Wind Nebulae (PWN) associations, thermal X-ray emission, halos and filamentary structures (\citealt{Seta_1998}, \citealt{Rho_1998}).\\
Recent single-dish imaging of W44 and IC443 was performed by \cite{Egron_2017} using the Sardinia Radio Telescope (SRT\footnote{www.srt.inaf.it}) data at 1.5 GHz and 7.0 GHz in the framework of the SRT astronomical validation (AV, \citealt{Prandoni_2017}) and the Early Science Program (ESP\footnote{www.srt.inaf.it/astronomers/early-science-program-FEB-2016/}).
For the first time, these observations provided deep single-dish imaging of W44 and IC443 at 7.0 GHz.  
By coupling 1.5$-$7.0 GHz maps, \cite{Egron_2017} also obtained spatially-resolved spectral measurements, which show a spread in the spectral slope distribution from flat (or slightly inverted) spectra corresponding to bright radio structures, to relatively steep spectra
($\alpha$ $\sim$ 0.7) in fainter radio regions of the SNRs. 
The observed spread in the
spectral slope distribution was attributed to distinct primary and  secondary (produced by hadronic interactions) electron
populations in the SNRs, resulting from different shock conditions and/or undergoing different cooling processes. 
Spectral studies also suggested a possible slight spectral steepening above $\sim$1 GHz for both sources, which could be related to a break in the distribution of primary or, more likely, secondary hadronic electrons  \citep{Egron_2017}.
However, further high-frequency/high-resolution imaging data were necessary in order to disentangle different theoretical models that could describe the high-frequency spectra of these SNRs. \\
With this aim, we carried out multi-feed imaging observations of W44 and IC443 at 21.4 GHz ($K$-band) with SRT, providing a morphological and spectral description of these complex SNRs at radio frequencies that are so far unexplored with a resolution below 1 arcmin.
Tycho SNR was also included in this observation project because of its simple morphology and the availability of extensive flux density measurements at up to high radio frequencies ($>$100 GHz). These measurements are useful for cross-checking our calibration and data analysis procedure. \\
In Section 2, we provide a detailed description of the observations of SNRs performed with SRT and the main steps of the data reduction. In Section 3, we present the resulting calibrated images of Tycho, W44 and IC443 at 21.4 GHz, and related flux density measurements. Section 4 is dedicated to the discussion of the different emission models that could explain the observed spectral features of these SNRs. In the final section, we summarise the main conclusions of our study.

\section{Observations and data reduction}

\subsection{The $K$-band setup with SRT}
\begin{figure}
\includegraphics[width=\columnwidth]{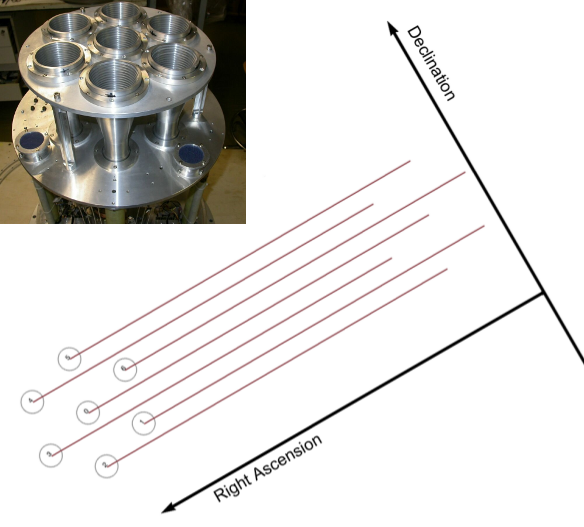}
    \caption{Schematic view of the ``On-the-Fly'' sky mapping using the SRT $K$-band 7-feed receiver (upper left) and adopting the ``Best Space Coverage configuration'' that maximises the sky co\-verage (Credits: Development of the Italian Single-dish COntrol System, DISCOS; \citealt{Orfei_2010})}
    \label{fig:multifeed}
\end{figure}

SRT is a 64-m single-dish radio telescope located in south Sardinia (Italy), and was designed for observations in the  0.3$-$116 GHz frequency range (\citealt{Ambrosini_2013}, \citealt{Bolli_2015}). It is characterised by an active surface (which consists of 1008 aluminium panels and of 1116 electromechanical actuators under computer control) on the primary mirror that corrects opto-mechanical deformations induced by gravity and temperature fluctuations, and improves spatial resolution at high frequencies (\citealt{Prandoni_2017}). 
SRT is presently equipped with three cryogenic dual-polarisation receivers placed at different focal positions (\citealt{Valente_2010}, \citealt{Navarrini_2016}, \citealt{Valente_2016}, \citealt{Navarrini_2017}): a 7-beam $K$-band receiver (18$-$26.5 GHz, Gregorian focus), a mono-feed $C$-band receiver (5.7$-$7.7 GHz, Beam Wave Guide focal position), and a coaxial dual-feed $L$/$P$ band receiver (0.305$-$0.41 GHz and 1.3$-$1.8 GHz, primary focus).
A detailed description of the $K$$-$band multi-beam receiver is given in \cite{Orfei_2010}.\\
During the SRT ESP observations in 2016, the spectral-polarimetric backend SARDARA (SArdinia
Roach2-based Digital Architecture for Radio Astronomy; \citealt{Melis_2018}) was configured in a mono-feed configuration.
The Total-Power (TP) backend, an analog double polarisation device, was available for all observing modes, including the multi-feed $K$-band configuration that we adopted for our observations. \\
As shown in Fig.\ref{fig:multifeed}, the $K$-band 7-feed receiver consists of a central feed and six lateral feeds arranged in a hexagonal configuration (the lateral/central feed separation in the focal plane is $\sim$100 mm, resulting in an angular separation in the sky of $\sim$2.3$\arcmin$). A mechanical rotator is used in the receiver to compensate for the Earth's rotation and maintain the parallactic angle, thus avoiding the astronomical field derotation caused by the SRT altitude-azimuth movement when tracking the sources in the sky.\\
Through the ``On-the-Fly'' (OTF) scanning mode, the antenna can scan the observed field along the axes of different reference frames at constant speed, which provides continuous data acquisition (the sampling time can be in the $1-1000$ ms range). Each single set of OTF scans achieved along two orthogonal axes forms an elementary map of the source.\\
The ``Best Space Coverage configuration'' (BSC; \citealt{Bolli_2015}) automatically rotates the dewar in order to best cover the scanned area, taking into
account the reference frame of the observation (Equatorial, Galactic and Horizontal).
For instance, if OTF scans are acquired in the Equatorial frame, the dewar is positioned so that the on-the-sky feed tracks are evenly-spaced in Right Ascension when scanning along Declination, and vice versa.
This specific geometry over the whole acquisition allows us to obtain, for each receiver scan, 7 simultaneous, equally-spaced (spacing $\sim$0.87$\arcmin$) OTF scans for both polarisation channels, which provide a full coverage of the scanned area (Fig.\ref{fig:multifeed}).
The seven-beam system increases the mapping speed of extended sources by a factor of seven when compared to a mono-feed. This fast mapping system is crucial for mitigating the effects of temporal atmospheric opacity variations on image quality \citep{Navarrini_2016}.

\subsection{SRT observations of SNRs}

SRT observations of SNRs Tycho, W44 and IC443 were performed in the framework of the ESP (S0009, PI A. Pellizzoni), between February and March 2016, at three different central frequencies: $L$-band (1.55 GHz), $C$-band (7.0 GHz) and $K$-band (21.4 GHz). In particular, the L and C-band observations and related data analysis techniques were described in detail by \cite{Egron_2017}. 
In this paper, we focus on  the continuum imaging of these SNRs using the multi-feed $K$-band receiver and the TP backend operating at a central frequency of 21.4 GHz, with a bandwidth of 1.2 GHz.
The corresponding beam size is $\sim$0.87 arcmin (half-power-beam-width; HPBW). \\
Our $K$-band observations were performed using the OTF/BSC scanning mode. 
Background emission and a system-related signal contribution forms the so-called ``baseline'', which is connected to the off-source OTF scan slope in its representation as signal-intensity vs. time plot. A precise subtraction of the baseline is very important in order to obtain an accurate flux density measurement of the observed source. From an observational point of view, this operation is guaranteed by requiring that at least 40$-$60\% of each scan be free from significant source contribution and radio-frequency interference (RFI) contamination (see also \citealt{Egron_2017}). We set the scan length of our observation according to the angular size of the targets (see Table \ref{tab:results_table}).
We adopted a scan speed of 4 arcmin/sec, which implies a duration for each scan of 10.4 sec for Tycho, 18 sec (RA) and 15 sec (Dec) for W44 and 22.5 sec for IC443.
Accounting for the beam size and a sampling time of 20 ms,
these observational parameters allow us to obtain about 10 measurement samples for each beam and scan.\\
Considering the pioneering nature of our observations with the new SRT multi-feed system, we imposed conservative observational requirements in order to obtain a complete map of the target for each individual feed and fully verify its imaging performances.
We scheduled an offset between two consecutive scans of the same map of $\sim$1/2 HPBW. Coupled with the 7 feed $K$-band receiver configuration (see Section 2.1), this allowed us to obtain about 140 samples/beam and an exposure time of 5.3 sec/beam for a full RA or Dec map.
This oversampling is fundamental to efficiently reject outlying measurements (due to RFI, individual feed anomalies, baseline subtraction issues), to perform an accurate evaluation of flux errors, and to optimise the final image accuracy.\\
Each scan is characterised by an additional dead/slew time that, for our setup, was about 30-40\% of the overall duration. The total duration of an observation resulting from the combination (averaging) of two consecutive RA and Dec maps is $\sim 2.5$ h  for W44, $\sim$ 4.2 h  for IC443, and $\sim$ 1 h for Tycho. \\
The observations were carried out in ``shared-risk mode'' and ideal atmospheric opacity conditions (high-quality $K$-band conditions are intended for $\tau$<0.1 Np) were not guaranteed. This had a negative impact on the quality of our preliminary SNR maps; this resulted in an increase in the rejection of flagged data. Indeed, the presence of an inhomogeneous cloud cover makes it difficult to identify (i.e. fit) both the baseline contribution and the weak signal of the source at this frequency, implying severe errors in a few data sets that were  prudentially discarded. \\
We performed observations of point-like flux density calibrators (3C286, 3C295, 3C147, 3C48 and NGC7027) through repeated OTF cross-scans in order to assess feed-dependent gain curves, and mapping to precisely estimate the beam shape.

\subsection{Multi-feed data reduction}

We performed the data analysis by using the ``Single-Dish Imager'' (SDI) software tool, which was tested during the SRT Commissioning and Scientific Validation (\citealt{Bolli_2015}, \citealt{Prandoni_2017}). SDI is designed to perform continuum and spectro-polarimetric multi-feed imaging starting from OTF scans, and is suitable for all SRT receivers/backends (see details in \citealt{Egron_2016}). The final SDI output are standard FITS images that can be used for further analysis with standard astronomy tools. \\
The first step of our data analysis procedure is the generation of the 14 individual feed images (separating left/right polarisation channels) related to the same target map. This is obtained through three subsequent steps: scan-by-scan baseline subtraction, RFI rejection, and production of preliminary calibrated images.
Throughout this procedure, we can assess the individual feed map quality in order to check the single-feed performance, and discard outlying samples as compared with the feed-averaged pixel value, before proceeding to the final merge (averaging) of all feed images. \\
The operations of baseline subtraction and RFI rejection are performed through an automatic process, and refined by a manual/interactive procedure.
The automatic baseline subtraction procedure provides a first rough estimate of the baseline parameters through a linear fit, and is applied scan-by-scan. 
For each scan, the angular coefficient and normalisation of the linear fit to the baseline are then refined through an iterative process aimed at optimising the number/percentage of scan samples that are within 1$\sigma$ level of the baseline fluctuation (see \citealt{Egron_2017} for details).
The automatic RFI rejection procedure identifies and removes the ``outlier'' samples presenting a count level above a standard deviation-based threshold (empirically chosen as a 5$\sigma$ level above the average of nearby samples taken within a distance of about 1/4 of the HPBW, in order to avoid discarding actual fluctuations
from the source). 
At the end of these filtering procedures related to the comparison of feed measurements, a significant fraction of ``outliers'' are identified and carefully discarded. Most rejected events are related to baseline subtraction anomalies due to atmospheric fluctuations, since the SRT $K$-band is relatively free from strong RFI contamination.
\begin{center}

\begin{table*}
\noindent
	\centering
	\caption{Flux densities of SNRs Tycho, W44 and IC443 for averaged maps and related image parameters obtained with SRT during the ESP in February-March 2016. The ``total duration''  refers to the total observation time for each target, including overheads. The exposure time associated with each target is of 21.2 sec/beam. A single map is intended as the combination of complete scans along RA and Dec directions.}
	\label{tab:results_table}
	\begin{tabular}{|l|c|c|c|c|c|c|c|c|r|} 
		\hline
	 & MJD & N.maps & Map Size & Total& \multicolumn{2}{c}{\qquad \qquad Extraction region: }  &Flux & rms\\
	      &      &  & ($^{\circ}$,$^{\circ}$) &duration & centroid coord (RA, Dec)& radius ($^{\circ}$) & (Jy) & (mJy/beam)\\
 \hline
\textbf{Tycho} &57442& 2 & 0.26$\times$0.26 & 1h30m & 0h25m19.4s, 64${^\circ}$0.8$\arcmin$ & 0.09 & 8.8  $\pm$ 0.9 & 23 \\
\hline \hline
\textbf{W44} & 57443, 57471  & 2 & 1.2$\times$1 & 5h9m & 18h56m05s, 01${^\circ}$21.6$\arcmin$ & 0.32 & 25 $\pm$ 3& 48\\
 \hline \hline
\textbf{IC443} &57442, 57468, 57470 & 2 & 1.5$\times$1.5 &  8h25m& 06h16m58s, 22${^\circ}$31.6$\arcmin$ & 0.5& 66 $\pm$ 7& 75\\
\hline 
	\end{tabular}
\end{table*}
\end{center}

\subsection{Data calibration and map production}

For the calibration of multi-feed data, we assumed that each feed (and polarisation channel)
is characterised by a specific efficiency that is mostly due to the different noise temperatures of the individual electronic chains (\citealt{Orfei_2010}).\\
We first evaluated a gain curve (as a function of the elevation) for the two polarisation channels of the central feed through OTF cross-scans on known flux density calibrators (3C286, 3C295, 3C147, 3C48 and NGC7027), and we adopted the \cite{Perley_2013} flux density scale.\\
A cross-scan is composed of four OTF scans ($0.5^{\circ}$ length and 1.25$\arcmin$/sec speed) that were alternately  performed along the RA and Dec directions. We performed the baseline subtraction and a Gaussian fit to the scanned data in order to measure the peak counts from the calibrator.
The gain curve was obtained through the calculation of a conversion factor, defined as the ratio between the observed counts from the calibrators and their expected flux density (counts/Jy). Our gain curves implicitly include the attenuation effect due to the atmospheric opacity (in the range 0.04$-$0.12 Np during the observations; \citealt{Buffa_2016},  \citealt{Buffa_2017}), which also affected the target observations. While the calibration measurements were taken temporally close to the target observations, the opacity conditions could in principle have been slightly different when we observed the calibrators and targets. We verified that the gain curve fluctuations due to a variable opacity during the observations were within the errors of the gain curve fit.
Calibration observations were performed for different elevations just before and after SNR mapping for each observing session, in order to account for atmospheric opacity fluctuations as a function of time.
As a further refinement, each target data sample was calibrated using the value calculated by averaging the calibration factors obtained within 6 hrs (or less in case of changing weather conditions) from the measurement, in order to guarantee comparable conditions for the target/calibrators observations.\\  
We checked that our gain curves for the central feed were consistent (within 8\%) with the one obtained during the AV of SRT (\citealt{Prandoni_2017}). Indeed the fluctuations due to the different atmospheric opacity conditions were within errors of the fit of our gain curves.
Gain curves (including opacity corrections) show a significant decrease in efficiency below $45^{\circ}$ of elevation ($\sim$40\% at $20^{\circ}$) due to a not perfectly optimised opto-mechanical configuration of SRT's active surface during the ESP.\\
We verified that gain curves of the lateral feeds only differ from the central feed by a normalisation factor.
In order to calculate this scaling factor for each feed/polarisation, we compared calibrator counts as seen by different feeds with the central feed counts.
We carried out calibrator maps ($0.2^{\circ}$$\times$$0.2^{\circ}$) during two sessions (25 February and 24 March 2016), by using the same central frequency, bandwidth, attenuation levels and OTF scan parameters adopted for the target observations.
Through 2D Gaussian fits of calibrator images for each feed, we calculated the peak count ratio between each lateral feed and the central feed. This scaling factor and the gain curves of the central feed allowed us to properly calibrate the lateral feeds.\\ 
The 14 data sets related to each feed/polarisation were individually mapped using the ARC tangent projection (pixels size equal to 1/2 HPBW) with respect to the world coordinate system, taking account of the actual feed positions.
All feed images were merged (averaged) together. Those strongly affected by high noise were removed, since they contributed to the overall image with a low signal-to-noise ratio.\\
A final filtering step removed measurements that were significant ``outliers'' with respect to the average (due to local gain fluctuation and/or badly removed baseline and/or residual RFI) for each pixel feed.
Considering all the SNR images, we discarded about 23\% of the data with a sensitive improvement of the final image quality. Despite this we still maintained about 100 good measurements per pixel.\\
In order to obtain brightness images (in units of Jy/sr), a precise evaluation of the beam solid angle is needed.
This was achieved through 2D Gaussian fits of the calibrator maps.
We obtained an averaged HPBW of 0.84$\arcmin$ $\pm$ 0.02$\arcmin$ at 21.4 GHz, which is only sightly dependent from the elevation in the observing range (15$^{\circ}$$-$80$^{\circ}$) for the target.
This value is consistent with the values obtained during both the technical commissioning and the scientific validation of SRT (\citealt{Bolli_2015}, \citealt{Prandoni_2017}).\\

\subsection{Error analysis}

Our final SNR images are characterised by a strong oversampling ($\sim 400$ good samples/pixel for a map obtained by merging two complete RA+Dec scans) that allows us to calculate the statistical errors on flux density measurements from the standard deviation of sample measurements for each pixel. The resulting statistical errors on the integrated flux densities of the targets are $<$0.5\%, which is a negligible value compared to the systematic errors.
The adopted flux densities for calibration sources \citep{Perley_2013} and their extrapolated errors to our observing frequencies imply errors on the flux density measurements of less than 1\%.\\
Our main source of uncertainties for flux density measurements is related to the errors associated with the gain curve estimation and the beam model applied to the calibration process.
We calculated the RMS errors from the fit of the gain curves obtained by our observation of calibrators at different elevations. The maximum RMS obtained from the different calibration sessions is $\sim$8\%.
The error on the estimate of the effective beam solid angle (and then on the brightness and integrated fluxes of the diffuse SNR targets) is $\sim$5\%, including slight fluctuations of the HPBW for different elevations and weather conditions.\\
Combining the above contributions, a conservative overall error on flux density measurements is thus $\sim$10\% (1$\sigma$ level). This value is comparable with the errors reported by \cite{Egron_2017} for SNR observations of W44 and IC443 with SRT in $L$-band ($\sim$3\%) and $C$-band ($\sim$5\%)\footnote{Smaller errors at low frequency observations mostly depend on the less important effect of (variable) atmospheric opacity on the measurements, especially at L-band.}.

\begin{figure*}
\includegraphics[width=8.5cm]{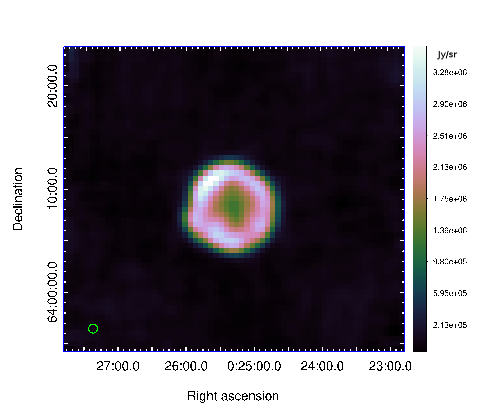}
\hspace{0.01cm}
\includegraphics[width=9.0cm]{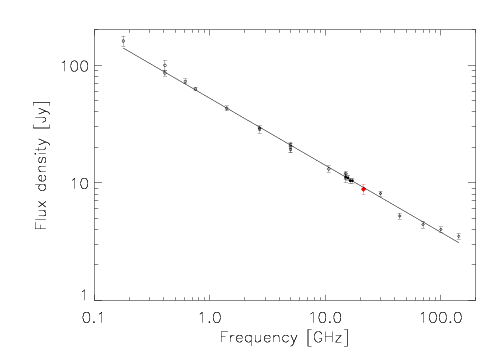}
    \caption{\textit{Left}: Map of Tycho SNR obtained with SRT at 21.4 GHz, taking into account a pixel size of 0.4$\arcmin$ (corresponding to $\sim$1/2 HPBW). The beam size is indicated by the green circle on the bottom left corner.  \textit{Right}: weighted least-squares fit applied to the Tycho SED for the synchrotron power-law model. The red and the black filled diamonds correspond respectively to the SRT point at 21.4 GHz and the $Planck$ data.}    
    \label{fig:SED 3C10}
\end{figure*}

\section{Results}

We present the total averaged maps of SNRs Tycho, W44, and IC443 in Fig.\ref{fig:SED 3C10} and \ref{W44_IC443}.
The data sets selected to obtain the final images and the imaging parameters are reported in Table\ref{tab:results_table} together with the image RMS and integrated flux density measurements. \\
The $K$-band observations performed with SRT are strongly dependent on the atmospheric conditions, i.e. the opacity and water vapor content (\citealt{Nasir_2011}, \citealt{Buffa_2017}).
Indeed, the observation of Tycho carried out on 24 February 2016 (MJD 57442) in optimal opacity conditions and in the absence of a cloud cover led to a high signal-to-noise ratio map that allowed us to perform an efficient baseline subtraction, which leads to a low RMS value (see Table\ref{tab:results_table}). \\
The high quality map of Tycho (Fig.\ref{fig:SED 3C10}) clearly shows the bright shell structure with its main filamentary structures, and the faint inner emission. In the case of W44 and IC443, the cloudy sky conditions and high humidity level attenuated the astronomical signal. This introduced noise in the measurements with a significant amount of rejected data and higher image RMS. Despite these weather conditions, the SRT map of W44 (Fig.\ref{W44_IC443}) shows the highly filamentary structure of the eastern  region. The brightest emission is along its boundary, and a short bright arc is visible in the western region where the SNR shock collides with a molecular cloud (\citealt{Mavromatakis_2003}, \citealt{Giacani_1997}).
In the W44 image, the Galactic plane is also clearly detected together with 
a nearby unidentified source located at (RA, Dec)=(18h57m04s, 1$^{\circ}$38$\arcmin$45$\arcsec$) that is spatially-resolved.
In the SRT map of IC443, we detect the main structures of the complex SNR morphology (Fig.\ref{W44_IC443}): the bright region corresponding to signatures of atomic/ionic shock in the northeastern part of the shell, the southern bright structure that includes the PWN region, and the main structures of the western halo region related to a
breakout portion of the SNR into a rarefied medium (\citealt{Lee_2008}).
\\
The SNR edges are clearly identified in our images when compared with other high-frequency data with lower resolution (e.g. \citealt{Planck_2016} and \citealt{Genova-Santos_2017}). 
Although the baseline-subtracted off-source pixels have zero mean flux in our maps, the rich image content requires a careful choice of the source extraction regions in order to properly calculate the integrated flux densities (see Table\ref{tab:results_table}). 
Slight variations of the above values do not significantly affect our flux density measurements since the background  in baseline-subtracted images is set to zero in regions that are free from unrelated source contamination. \\ 
We also estimated the flux densities related to the bulk of the bright radio emission located 
in the eastern boundary of W44 (the largest blue region in Fig.\ref{IC443_regions}) and in the northeastern region of IC443 (indicated by green contours in Fig.\ref{IC443_regions}), obtaining 13 $\pm$ 1 Jy and 14 $\pm$ 1 Jy, respectively.
The flux density associated with the unidentified  source  placed  east of W44 is 1.2 $\pm$ 0.1 Jy.
The map of W44 also includes a section of the Galactic plane contributing with an integrated flux density of 8.4 $\pm$ 0.8 Jy, which is calculated on a box centred at ($\alpha,\delta$)=(18h54m20.618s, +1$^{\circ}$26$\arcmin$28.23$\arcsec$) and an area of $1^{\circ}\times0.3^{\circ}$ (highlighted in Fig.\ref{IC443_regions}).
Lower resolution images could include these significant contributions to the apparent total flux of W44. 
\section{Discussion}
Sensitive high-resolution radio observations at up to $\sim$18 GHz are reported in the literature only for Tycho, while for SNRs W44 and IC443, the highest reported radio frequencies are at up to 5$-$10 GHz. $Planck$ observations of W44 and IC443 were performed at up to 70 GHz and 857 GHz, respectively, with an angular resolution in the $5\arcmin-30\arcmin$ range (\citealt{Planck_2016}).   
SRT $K$-band observations of W44 and IC443 provided the first ever obtained images above 10 GHz with sub-arcmin resolution, which is enough to obtain accurate continuum flux density measurements and spectra. We performed the weighted least-square fit of Tycho, W44 and IC443 radio spectra using the MPFIT\footnote{http://purl.com/net/mpfit} (\citealt{Markwardt_2009}) package. 

\subsection{ Tycho }

Observed by Tycho Brahe in 1572, Tycho SNR (also named 3C10) resulted from a Type Ia Supernova (SN) explosion.
It is one of the youngest SNRs in the Galaxy, and is expanding into the ISM without a strong radial density gradient. The average expansion rate is approximately  0.1\% yr$^{-1}$, as expected for a SNR in transition from a rapid expansion to the Sedov adiabatic phase (\citealt{Reynoso_1997}). \\
At radio wavelengths, Tycho exhibits a well defined loop structure with an angular diameter of 8$\arcmin$ and a total continuum flux density of 56 Jy at 1 GHz.
The radio brightness distribution of Tycho indicates a thin outer rim where both shock compression and particle acceleration occur, and a thicker
main shell in which instabilities and turbulent amplification may be  responsible for the emission (\citealt{Dickel_1991}). \\
The integrated spectral index calculated in the 0.178$-$17.1 GHz frequency range is $\alpha$ = 0.58 $\pm$ 0.02 \citep{Sun_2011}.
A slightly concave up model (hardening to higher frequencies) was proposed to describe the integrated spectrum of Tycho (\citealt{Reynolds_1992}). 
 Indeed, a spectral flattening (thermal Bremsstrahlung component) at high frequencies combined with an integrated spectral index in the range 0.3$-$0.7, are considered  to be typical characteristics of ``shell'' type remnants of which Tycho is thought to be a member (\citealt{Hurley-Walker_2009}).\\
Spatially resolved spectral-slope studies of Tycho showed evidence for a significant spectral steepening close to the SNR centre. This result suggested a high-energy electron distribution in the thinner shell, where the dense thermal material has been swept up from the interstellar medium (\citealt{Duin_1975}). A following study of the spectral index distribution contradicted this result,  revealing no significant spatial variations \citep{Klein_1979}.
More recent sensitive low-frequency VLA observations (0.3$-$1.4 GHz, \citealt{Katz-Stone_2000}) identified different radio structures within Tycho, allowing for a spatial study of the spectral index across the entire SNR. Brighter filaments located in the outer regions  display flatter spectral indices than the fainter ones. This may be due to competing mechanisms of SNR blast wave-ambient medium interactions and internal inhomogeneities of the magnetic field within the remnant (\citealt{Katz-Stone_2000}). \\   
We analysed the radio spectrum of the SNR Tycho by considering all of the flux density measurements available in the literature for the 0.038$-$143 GHz frequency range.
Flux density values at up to 17 GHz were taken from
Tables 1 of \cite{Klein_1979} and of  \cite{Sun_2011}. For the high radio frequencies, we used the measurements obtained between 15 and 17.1 GHz with the \textit{Arcminute Microkelvin Imager Small Array} (AMI SA) given in Table 4 by \cite{Hurley-Walker_2009}, our measurements obtained at 21.4 GHz with SRT, and the $Planck$ measurements carried out at up to 140 GHz, which are reported in Table 3 of the \cite{Planck_2016}. 
The overall radio SED for Tycho is displayed in Fig.\ref{fig:SED 3C10}. The SRT value at 21.4 GHz (represented with a filled red diamond) perfectly matches the trend suggested by the other data without any apparent spectral variation.  \\
The data are fitted ($\chi^{2}/dof$= 1.3) by a synchrotron emission model represented by a simple power law function (with a normalisation constant of 52.3 $\pm$ 0.7 Jy that represents the flux density at 1 GHz), which rules out spectral flattening models.
The resulting spectral index $\alpha=0.58 \pm0.01$ is consistent with the value obtained by \cite{Sun_2011}.
A spectral steepening is also unsupported by our points, which confirms the trend suggested by $Planck$ data.
This is consistent with the predictions based on gamma-ray observations of Tycho carried out with VERITAS, and interpreted by \cite{Acciari_2011} through leptonic and hadronic models, which suggest a steepening of the radio spectrum at much higher frequencies (above $\sim 10^{14}$ Hz). Both these models provide a reasonable fit to the gamma-ray data and imply a magnetic field of $\sim$80 $\mu$G, which is possibly  interpreted  as  evidence  for magnetic field amplification (\citealt{Morlino_2012}). \\

\begin{figure*}
\includegraphics[width=9.cm]{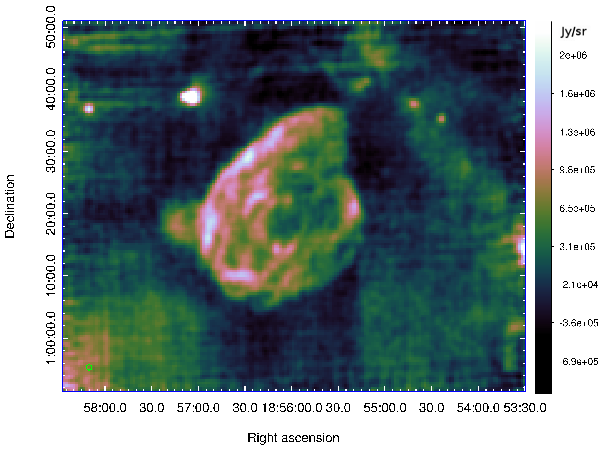}
\label{W44}
\hspace{0.01cm}
\includegraphics[width=8.5cm]{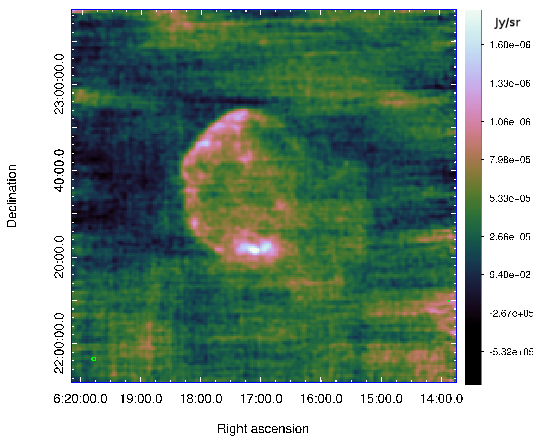}
\caption{Map of SNRs W44 (left) and IC443 (right) obtained with SRT at 21.4 GHz. The pixel size is 0.4$\arcmin$. The beam size is indicated by the green circle on the bottom left corner of the maps. Sky opacity-related artefacts, particularly evident in the IC443 image, minimally affect the flux density estimates. A portion of the Galactic Plane is visible in the top-right corner of the W44 image (see also Fig.\ref{IC443_regions} left).} 
    \label{W44_IC443}
\end{figure*}

\subsection{W44}

\begin{figure*}
%
%
%
%
\includegraphics[width=8.7cm]{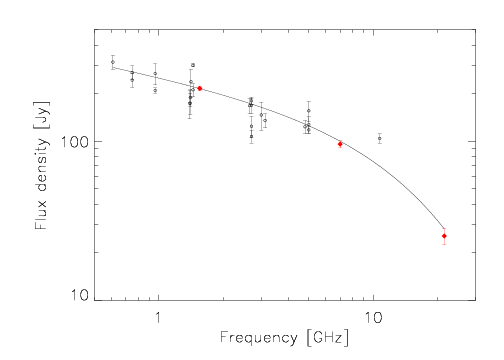}
\label{W44_onicSRT_p}
\hspace{0.001cm}
\includegraphics[width=8.7cm]{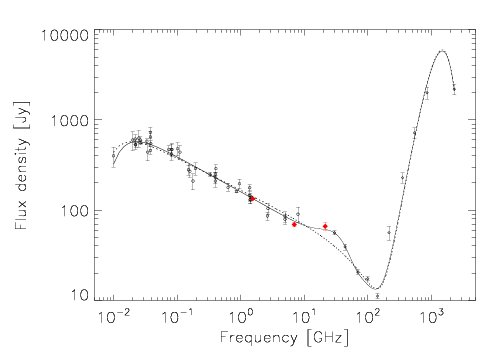}\\
\small (a) W44 \qquad \qquad \qquad \qquad \qquad \qquad \qquad \qquad \qquad \qquad \qquad \qquad \qquad\small (b) IC443
 \caption{a) Weighted least-squares fit applied to the W44 SED for the synchrotron power law with an exponential cut-off model. The red diamonds correspond to the SRT flux density measurements at 1.55, 7.0 and 21.4 GHz. b) SED of IC443 from 0.408 to 857 GHz. Open circles and diamonds represent radio data from the literature at up to 8 GHz and the Planck data, respectively. SRT measurements are indicated
with filled red diamonds. The solid and dashed lines represent the
weighted least-square fit considering the spinning dust emission
and without it, respectively.}
    \label{W44_IC443_SED}
\end{figure*}

\begin{figure*}
%
%
%
%
\includegraphics[width=9.cm]{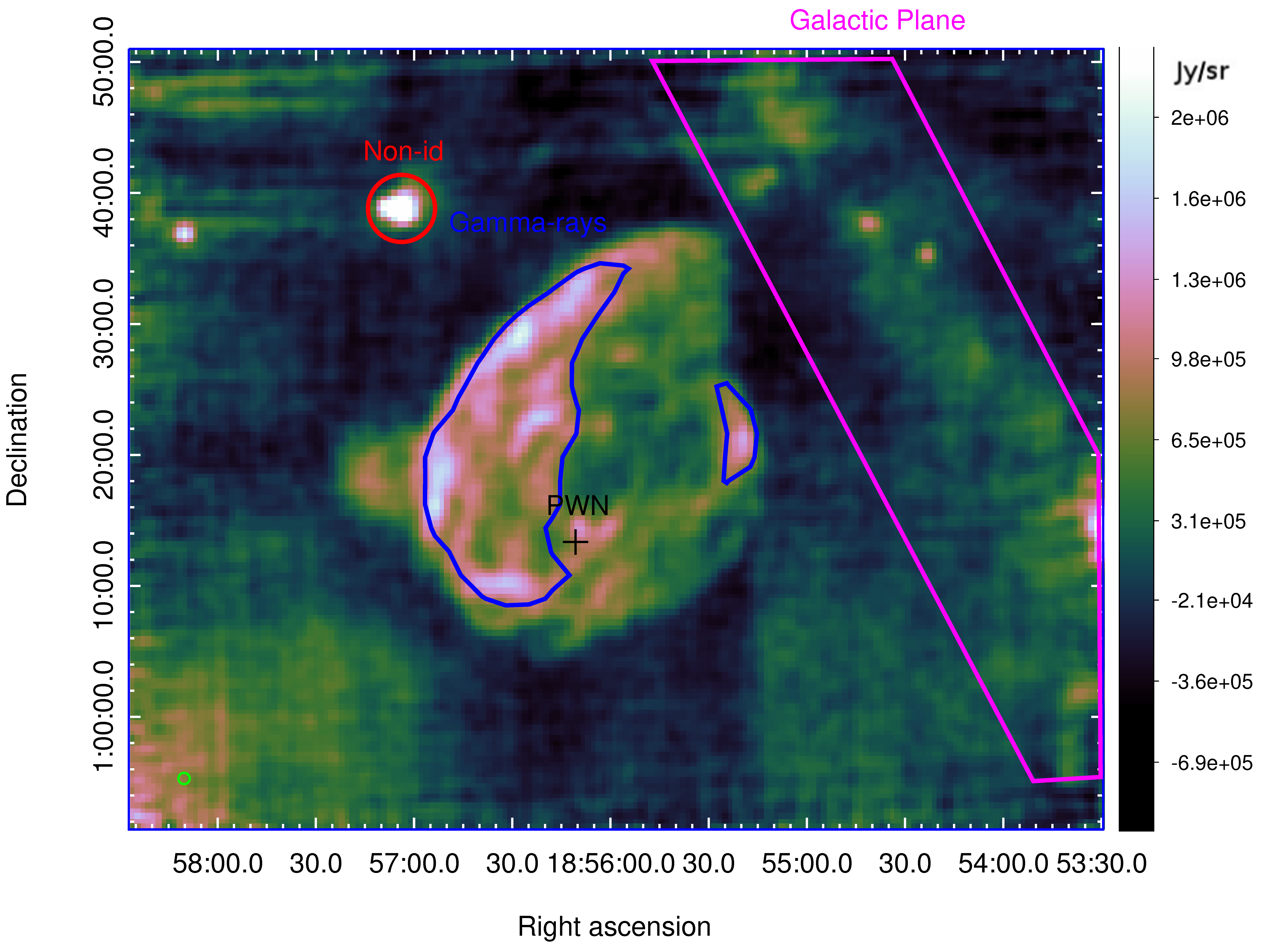}
\hspace{0.01cm}
\includegraphics[width=8.5cm]{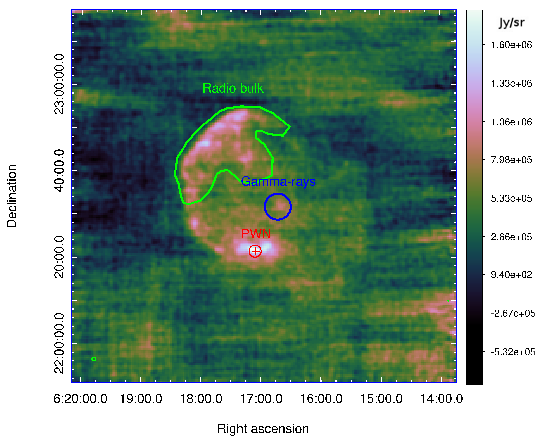}
\caption{\textit{Left}: continuum map of SNR W44 at 21.4 GHz. The blue regions indicate the  gamma-ray emission seen by AGILE and Fermi-LAT (\citealt{Giuliani_2011}, \citealt{Abdo_2010}). The red circle refers to the bright non-identified source described in the text. The black cross shows the position of the PWN. The magenta region indicates the Galactic plane. \textit{Right}: Continuum map of SNR IC443 at 21.4 GHz. The red cross and circle indicate the position of the compact source CXOU J061705.3+222127 and the associated PWN, respectively. The blue circle indicates the  gamma-ray emission seen with MAGIC \citep{Albert_2007}. The green region shows the bulk of the radio emission. } \label{IC443_regions}
\end{figure*}

W44 is a middle-aged SNR with an estimated age of $\sim$20,000 yrs and an angular dimension of $\sim$30$\arcmin$ (\citealt{Wolszczan_1991}, \citealt{Smith_1985}).
W44 is located in a crowded region of the Galactic plane at a distance of $\sim$ 3 kpc, in the inner part of the W48 molecular cloud complex \citep{Onic_2015}.
It has been shown that W44 represents a typical case of a SNR-molecular cloud interacting system. In the radio band, this SNR is characterised by an asymmetric limb brightened shell. The brightest emission is located along the eastern boundary (RA, Dec = 18h56m50s, 01$^{\circ}$17$\arcmin$), where the SNR interacts with dense molecular clouds (\citealt{Seta_2004}, \citealt{Reach_2005}). In the western region, a short bright arc is visible (RA, Dec= 18h55m20s, 01$^{\circ}$22$\arcmin$), and is related to the interaction between the SNR shock and a molecular cloud. Bright filaments located in the eastern inner region (also visible in Fig.\ref{W44_IC443}) result from radiative shocks encountering the clouds or sheets of dense gas (\citealt{Jones_1993}). \\ 
The  $K$-band flux density measured for W44 is far below the extrapolation of the spectrum obtained through low-frequency SRT measurements \citep{Egron_2017}.
We estimate a spectral index of $\alpha=1.26\pm0.09$ using the integrated flux density measurements carried out with SRT at 7.0 GHz and 21.4 GHz. 
A comparison between this result and that obtained by \cite{Egron_2017} $\alpha=0.55\pm0.03$ (1.55$-$7.2 GHz) suggests a significant spectral index steepening at high frequencies.\\
In Fig.\ref{W44_IC443_SED} (a), we show a comprehensive picture of the  SED of W44  in the frequency range between 610 MHz and 21.4 GHz, which was obtained by using the flux density measurements from Table 2 by \cite{Castelletti_2011}\footnote{Only data corrected to the scale of \cite{Baars_1977} are taken into account} and those obtained with SRT at 1.5 GHz, 7.0 GHz (\citealt{Egron_2017}) and 21.4 GHz (this work). 
Further high-frequency measurements between 30 GHz and 70 GHz were carried out by $Planck$ with an angular resolution from 31$\arcmin$ to 5$\arcmin$ (\citealt{Planck_2016}). Since the resolution of \textit{Planck's Low  Frequency Instrument} (LFI) makes it difficult to obtain sensitive flux density measurements of a source located in a very crowded portion of the Galactic plane, we decided to exclude $Planck$ data from our study. Similar arguments may apply to  low-resolution QUIJOTE data at 10$-$20 GHz \citep{Genova-Santos_2017}. \\
Taking into account the significant scatter of flux density measurements and the low availability of high-frequency measurements, we modelled the integrated spectrum  of W44 using a simple synchrotron model with an exponential cutoff expressed by $S(\nu)= K \left(\frac{\nu}{\nu_0}\right)^{-\alpha} e^{-\frac{\nu}{\nu_0}}$,
where K is a normalisation constant and $\nu_0$ is the cutoff frequency.\\
The weighted least-squared fit (Fig.\ref{W44_IC443_SED}, a) gives a spectral index $\alpha=0.24\pm0.04$, a cutoff frequency of $\nu_0=15\pm2$ GHz and K = 267 $\pm$ 5 Jy. \\ 
This result is consistent with the synchrotron break at frequencies $\gtrsim$10 GHz predicted by the leptonic model based on gamma-ray observations (\citealt{Ackermann_2013}).\\
An electron of energy $E$ accelerated in a magnetic field B, radiates its peak power at a frequency $\nu$ so that $E= 14.7(\nu_{GHz}/B_ {\mu G})^{\frac{1}{2}} GeV $ \citep{Reynolds_2008}.
Assuming a magnetic field in the range 18$-$90 $\mu$G 
(\citealt{Castelletti_2011}, \citealt{Ackermann_2013}) and the derived cutoff frequency of 15 GHz, we obtain a maximum energy of the electron distribution in the $\sim$6$-$13 GeV range. This result is in agreement with the particle maximum energy of $\sim$10 GeV estimated from the gamma-ray data (\citealt{Uchiyama_2012}, \citealt{Ackermann_2013}).
The magnetic field interval reported above is typically in agreement with leptonic-dominated emission scenarios and could also be simply explained by the compression theory of the ISM magnetic field (about 5 $\mu$G, \citealt{Wielebinski_2005}) by a factor of  3 to 4.\\
In order to properly investigate the co-spatial radio-gamma emitting regions, we restricted our analysis to the regions shown with blue contours in Fig.\ref{IC443_regions} that correspond
 to the gamma-ray emissions detected with AGILE and Fermi-LAT (\citealt{Giuliani_2011}, \citealt{Abdo_2010}). The eastern gamma-ray region was considered by \cite{Cardillo_2016} to study the radio and gamma-ray spectra from W44 in terms of re-acceleration and compression of Galactic CRs.
Using the flux density measurements obtained for this region from the SRT maps in the L, C \citep{Egron_2017} and K-band, we found a cutoff frequency of 16 $\pm$ 1 GHz, which is consistent and slightly higher than the corresponding value obtained from the total integrated flux for W44.
Coupling this result with the locally enhanced magnetic field of 1.4 mG deduced by \cite{Cardillo_2016} for this region, we obtained a maximum electron energy of 1.6 GeV. \\
This lower energy cutoff with respect to the $\sim$10 GeV limit obtained from standard magnetic field values, could suggest the presence of a spectrum dominated by secondary electrons produced from pion decay following proton-proton scattering. 
Indeed, as reported by \cite{Cardillo_2016}, electrons with a maximum energy $\sim$1$-$2 GeV could result in crushed clouds regions from protons with a cutoff energy at $\sim$ 10 GeV. \\
We studied the spectral indices associated with all the different peculiar regions of W44 shown in Fig.\ref{IC443_regions} by using the flux density measurements carried out with SRT at 7.0 GHz \citep{Egron_2017} and 21.4 GHz (this work). In our analysis, we also included a region centred in RA, Dec =18h55m49.9s, 1$^{\circ}$20$\arcmin$00.8$\arcsec$ with radius 4.5$\arcmin$, which is representative of the fainter central emission.
We obtained $\alpha = 1.0 \pm 0.1$ for the eastern boundary,  $\alpha$ = 0.9 $\pm$ 0.1 for the western arc and $\alpha$ = 1.4 $\pm$ 0.1 for the central region.
The comparison of these spectral indices with those calculated by \cite{Egron_2017} in the 1.55$-$7.0 GHz range, confirms the steepening observed in the integrated spectrum of W44 for all of the considered SNR sub-regions.\\
On the other hand, this result suggests a spectral index spread ranging from relatively flatter spectra, which is associated with bright radio limbs and filaments, to  steeper spectra in fainter radio regions. The same region-dependent scatter in spectral slope distribution was highlighted by spatially-resolved spectral measurements in the 1.5$-$7.0 GHz frequency range (\citealt{Egron_2017}), and might suggest the existence of distinct electron populations subject to different shock conditions and/or undergoing
different cooling processes.
Furthermore, we observe a correspondence between bright-flat spectrum regions and gamma-ray emission that could represent a signature of secondary electron populations produced by hadronic interactions in regions where the SNR
shock collides with dense molecular clouds \citep{Cardillo_2016}.
An unidentified source is spatially resolved by SRT at 21.4 GHz in the north-east direction, with an associated flux density of 1.2 $\pm$ 0.1 Jy.
Very Large Array (VLA) radio images at 74 and 324 MHz \citep{Castelletti_2007} and \textit{Spitzer Space Telescope} infrared observations at 24 $\mu$m and 8 $\mu$m \citep{Reach_2006} detected this source that could be associated with the G035.040-00.510 H\textsc{ii} region (WISE Catalog of Galactic H\textsc{ii} regions\footnote{astro.phys.wvu.edu/wise/}). This source was also detected as a point-like source with SRT at 7.0 GHz \citep{Egron_2017} with a flux density of 0.94 $\pm$ 0.05 Jy. The associated spectral index obtained combining the SRT flux density measurements at 7.0 GHz and 21.4 GHz is $\alpha$ = $-$0.2 $\pm$ 0.1 (implicitly assuming non-variability of the source flux).\\

\subsection{IC443}
 
IC443 is a large (45$\arcmin$ size) evolved SNR formed nearly 4,000 yrs ago \citep{Troja_2008}, and located at a distance of $\sim$1.5 kpc \citep{Petre_1988}. 
The presence of both diluted and dense gaseous environments surrounding the SN blast wave and the interaction of the SN shock with several atomic and molecular clouds, makes IC443 a very interesting system with which to assess the scenario of CRs acceleration in SNRs (\citealt{Lee_2008}, \citealt{Tavani_2010}, \citealt{Onic_2017}).\\
Its location close to the Galactic anti-centre, which is relatively isolated from the confusing contamination by unrelated Galactic plane emission, also makes this SNR a suitable target for low resolution radio and gamma-ray observations (\citealt{Planck_2016}, \citealt{Tavani_2010}).\\ 
In the radio domain, IC443 appears to be composed of two nearly concentric shells of synchrotron emission. The shell located in the eastern part of the SNR includes the bulk of the radio emission and is connected in its western side to the weaker second shell (halo), which is possibly related to a breakout portion of the SNR into a rarefied medium \citep{Lee_2008}. The bright-southern boundary is associated with the interaction between the shock and a mostly molecular medium \citep{Burton_1988}. A bright cometary-like PWN is located in the same region, and is associated with the compact source  CXOU J061705.3+222127. The PWN, $\sim$2$\arcmin \times$1.5$\arcmin$ in size, is characterised by the synchrotron emission from relativistic particles generated by the pulsar that interacts with
the SNR shell \citep{Swartz_2015}. \\ 
Integrated spectral studies performed on IC443 showed a turnover at the lowest radio frequencies (below $\sim$ 30 MHz) that is attributed to thermal absorption processes
 at the SNR site \citep{Castelletti_2011}. \\
From spatially-resolved spectral index studies, \cite{Castelletti_2011} highlighted a correspondence between the bright/flattest spectrum eastern radio region and the near-infrared ionic lines.
Considering this scenario, the thermal absorption of electrons would be the main mechanism responsible for the spectral flattening in this region \citep{Rho_2001}.
\cite{Egron_2017} showed a spectral index distribution of IC443 whose spread could instead be due to an intrinsic variety in the primary and secondary electron spectra (spectral slopes and breaks)
produced in shocks that are located in different SNR/PWN environments, meaning that several region-dependent electron populations are
present.\\
The possibility of a significant production of thermal Bremsstrahlung radiation at radio continuum frequencies was proposed by \cite{Onic_2012} to describe the IC443 spectrum at high radio frequencies (at up to $\sim$ 10 GHz) with a ``concave up'' model. \\
We performed a weighted fit of the SRT data at 1.5 GHz, 7.0 GHz \citep{Egron_2017} and 21.4 GHz (this work) with a simple power-law model, and we obtained a spectral index of  $\alpha$ = 0.38 $\pm$ 0.03.
This result is 1.3$\sigma$ from the integrated spectral index of $\alpha$ = 0.46 $\pm$ 0.03 obtained by \cite{Egron_2017} in the 1.55$-$7.0 GHz interval. 
In the same work, \cite{Egron_2017}
highlighted a slight steepening of the IC443 spectrum ($\Delta\alpha$$\sim$0.1) around $\sim$1 GHz, as suggested by the comparison between the SRT results and integrated spectral indices obtained from the literature data in the  0.02$-$1.0 GHz frequency range (\citealt{Egron_2017}, \citealt{Castelletti_2011}). 
By coupling these spectral features with our measurement at 21.4 GHz, we could support the concave-up curvature of the radio spectrum  firstly suggested by \cite{Onic_2012}.\\
However, a radio-microwave study  performed by the \cite{Planck_2016} showed no indication of thermal Bremsstrahlung emission and suggested a combination of synchrotron and dust emission to explain the integrated spectrum of IC443 in the 1$-$857 GHz frequency range.
$Planck$ data suggested the presence of a spectral bump in the 20$-$70 GHz range in contrast with the hypothesis of a concave-up spectrum.\\
Recent work \citep{Onic_2017}  that coupled $Planck$ data with microwave observations from space telescopes such as the \textit{Wilkinson Microwave Anisotropy Probe} (WMAP), again showed a synchrotron component at up to $\sim$70 GHz with an additional contribution from Anomalous Microwave Emission (AME) due to the spinning dust process that could explain the observed bump.
The spinning dust emission is based on a mechanism whereby the dust asymmetric grains with a non-zero electric dipole moment may be spinning due to the interaction with the ISM and radiation field, and thus radiate electromagnetic waves due to the rotation of their electric dipole moment \citep{Ali_2009}. 
The study of Galactic clouds performed by the \cite{Planck_2014}  suggested that the AME could come from molecular clouds dust or photodissociation regions (PDR). 
The higher-frequency emission observed in IC443 above $\sim$150 GHz could instead be attributed to thermal dust emission form the dust grains that survived the shock \citep{Planck_2016}.\\
In Fig.\ref{W44_IC443_SED} (b), we present the spectral model of IC443 proposed by \cite{Onic_2017}, based on least-squares fits ($\chi^{2}/dof$= 2.3) of integrated flux density measurements from Table 1 of \cite{Castelletti_2011}, Table 1 of \cite{Gao_2011}, Table 3 of \cite{Reich_2003} and microwave measurements from Table 3 of the \cite{Planck_2016}.
The resulting SED is modeled  by a synchrotron emission with an exponential cutoff at 148 $\pm$ 23 GHz and a normalisation constant of 161 $\pm$ 3 Jy, spinning dust emission (peaked at the frequency 28 $\pm$ 3 GHz) represented by the expression shown in Equation (5) by \cite{Onic_2017} and a high-frequency ($>$ 140 GHz) thermal dust emission component.
SRT data at 1.5 GHz, 7.0 GHz and 21.4 GHz are consistent with the model that includes the spinning dust emission. 
In particular, our flux density measurement at 21.4 GHz independently supports the observed spectral bump, and confirms the trend suggested by the low-resolution $Planck$ measurements at 30 GHz.\\
The possibility of a significant spinning dust emission from a SNR was introduced for the first time by \cite{Scaife_2007} in a study of the SNR 3C96, which was not later confirmed. Later on, there was a possible detection of spinning dust emission from W44 (\citealt{Genova-Santos_2017}). However, this result is strongly affected by unresolved emission contributions from the Galactic Plane. Within this framework, our result on IC443 would be the first confirmation of spinning dust emission in a SNR.
\\
We coupled our map of IC443 at 21.4 GHz with that obtained with SRT at 7.0 GHz \citep{Egron_2017} in order to study the spectral indices associated with different peculiar regions across the SNR.  As displayed in Fig. \ref{IC443_regions}, we analysed the northeastern shell region related to the bulk of the radio emission, the region associated with the PWN, and those coincident with the gamma-ray emission as seen with MAGIC \citep{Albert_2007}.\\
The spectral index associated with the brightest part of the eastern shell is steeper ($\alpha=$ 0.35 $\pm$ 0.1) than the averaged value obtained in the same frequency range ($\alpha=$ 0.04 $\pm$ 0.1). This result indicates that the spectral bump related to the spinning dust emission is not evident in the eastern region.
A spectral flattening related to the brightest SNR regions is  evident when considering the region with an extraction radius of 1.4$\arcmin$, 
which includes its associated PWN described at lower radio frequencies by \cite{Castelletti_2011}. We obtain a flux density of 0.53 $\pm$ 0.05 Jy and a flat spectral index $\alpha$ = $-$0.05 $\pm$ 0.09 (7.0$-$21.4 GHz), in agreement with that reported in previous studies \citep{Castelletti_2011},
and probably associated with optically thin PWN emission.
In the bright southern region  that includes the PWN and where the SNR interacts with a molecular cloud, the spinning dust emission  could further contribute to the observed spectral flattening.\\
The VHE $\gamma$-ray source associated with IC443, such as seen by MAGIC \citep{Albert_2007}, is located at (RA, Dec)=(06h16m43s , +22$^{\circ}$31$\arcmin$48$\arcsec$), and occupies approximately a circular region with radius $\sim$ 0.05$^{\circ}$. Previous studies have highlighted the morphological coincidence between the TeV emission and a molecular gas distribution \citep{Castelletti_2011}.
As shown in Fig.\ref{IC443_regions}, our 21.4 GHz image reveals a bright radio structure corresponding to the TeV source, with a flat spectral index of 
$\alpha$ = 0.03 $\pm$ 0.09 (7.0$-$21.4 GHz). \\
The correlation between the TeV emission and bright radio/flat (or slightly inverted) spectral regions could be related to an increase in hadron emission associated with a significant secondary electrons injection (\citealt{Lee_2015}, \citealt{Cardillo_2016},  \citealt{Egron_2017}). 
The co-spatial presence of a molecular cloud interacting with the SNR
could imply the presence of a CR acceleration by the shock in a region characterised by strong post-shock densities and an enhanced magnetic field, as result of the fast decrease of the shock speed in a dense medium and an increasing role of the radiative
losses \citep{Petruk_2016}.

\section{Conclusions}

The $K$-band 7-feed receiver available at SRT is well suited to study extended sources like SNRs. It provides smart and sensitive   mapping at a sub-arcmin resolution, reducing the effects of temporal atmospheric opacity variations that are particularly severe at high-radio frequencies \citep{Navarrini_2016}.\\  
In their early phase, young SNRs such as Tycho generally do not show a spectral break in the radio band due to very efficient acceleration conditions \citep{Urosevic_2014}. Our flux density measurement on Tycho obtained with SRT at 21.4 GHz confirms the non-thermal synchrotron as the dominant emission process, and this rules out any spectral curvature up to high radio frequencies.\\
On the other hand, a spectral index steepening is  expected at radio frequencies in evolved SNRs wherein the particle acceleration mechanisms are not efficient (\citealt{Ackermann_2013}, \citealt{Urosevic_2014}). Although comparable in age, our observations show that, surprisingly, W44 and IC443 present different cutoff energies in their synchrotron spectra.\\
For the first time, we observed  a synchrotron spectral break in SNR W44 at a frequency of 15 $\pm$ 2 GHz. This result provides a direct estimate for the maximum energy of accelerated CR electrons  in the 6$-$13 GeV range, which is consistent with indirect evidence from gamma-ray observations (\citealt{Ackermann_2013}, \citealt{Cardillo_2014}, \citealt{Cardillo_2016}).\\
In the case of IC443, \cite{Onic_2017} estimated a synchrotron cutoff frequency (148 $\pm$ 23 GHz) of an order of magnitude higher than what we found for W44, with the concurring presence of a significant dust emission bump at 20$-$70 GHz, which is confirmed by our observations at 21.4 GHz.  
We do not rule out the possibility that a dust emission component could also be present in the W44 SED \citep{Genova-Santos_2017}. This hypothesis could be confirmed by further high-resolution measurements in the crowded field of W44 in the 10$-$100 GHz range.\\
Synchrotron cutoff differences between W44 and IC443 could be related to different primary and secondary electron energy distributions and/or magnetic fields, according to local shock conditions arising from different interacting structures in the ISM \citep{Egron_2017}. 
Deeper SNR images could be provided by SRT through  the recently implemented spectro-polarimetric backend SARDARA \citep{Melis_2018}, which, coupled with the multi-feed receivers at up to 50 GHz (a Q-band receiver, in the frequency range of 33$-$50 GHz, is currently being implemented  for SRT, \citealt{Navarrini_2016}), is perfectly suited to fully constrain synchrotron spectral breaks in SNR SEDs that are tightly related to Galactic CR acceleration.

\section*{ACKNOWLEDGEMENTS}
The Sardinia Radio Telescope is funded by the Department of University and Research (MIUR), the Italian Space Agency (ASI), and the Autonomous Region of Sardinia (RAS), and is operated as a National Facility by the National Institute for Astrophysics (INAF). We acknowledge the very useful comments of the anonymous referee.
S. Loru gratefully acknowledges the University of Cagliari and INAF for the financial support of her PhD scholarship. M. Pilia was supported by the Sardinia Regional Government through the project ``Development of a Software Tool for the Study of Pulsars from Radio to Gamma-rays using Multi-mission Data'' (CRP-25476).
We are very grateful to R. G{\'e}nova-Santos for the useful discussion about Planck and QUIJOTE data.




\bibliographystyle{mnras}
\bibliography{example} 








\bsp	
\label{lastpage}
\end{document}